\documentclass[journal]{IEEEtran}
\usepackage{cite}

\ifCLASSINFOpdf
\else
\fi
\usepackage{amsmath}
\usepackage{amssymb}

\usepackage{url}

\usepackage{txfonts}

\usepackage{CJKutf8}
\usepackage{graphicx}

\begin{document}
%
%
%

\title{Anisotropic non-Hermitian skin effect in a two-dimensional Lieb photonic crystal}

%

\author{Zhi-Kang Xiong, ~T. Wen,~
        Y. Liu (刘泱杰),~\IEEEmembership{Member,~IEEE, }
        H. Lin,~\IEEEmembership{Member,~IEEE,} and~Bin~Zhou
\thanks{C. Z., T. W., Y. L., and B. Z. are with Department of Physics, School of Physics, Hubei University, Wuhan 430062, P. R. China. B. Z. is also with Wuhan Institute of Quantum Technology, Wuhan 430206, P. R. China. Corresponding authors' (Y. L., and B. Z.) email addresses: yangjie@hubu.edu.cn, and binzhou@hubu.edu.cn}%
\thanks{H. L. is with College of Physical Science and Technology, Central China Normal University, Wuhan 430079, P. R. China.}

\thanks{Manuscript submitted on 1st Mar., revised 15-17th May 2026. This work has been submitted to the IEEE for possible publication of Special Issue on Emerging Topics for Nanophotonics, Metamaterials and Plasmonics, 2026. Copyright may be transferred without notice, after which this version may no longer be accessible. }}

%
%

\markboth{IEEE Journal of Selected Topics in Quantum Electronics, Special issue, May~2026}%
{Xiong \MakeLowercase{\textit{et al.}}}
%



\begin{CJK}{UTF8}{gbsn}
\maketitle

\end{CJK}

\begin{abstract}
	In this contribution paper, we construct a two-dimensional non-Hermitian (NH) photonic crystal (PhC) to prototype its anisotropic non-Hermitian skin effect (NHSE) for experimental proposal. Based on the tight-binding model for Lieb lattice with NH coupling, a nontrivial spectral winding number is pinpointed for certain eigenstates, which translates to geometry-dependent skin modes with tilt boundaries. For ease of implementation, complex refractive indices are employed for the Lieb unit cell of PhC to emulate the NH coupling. Validated by full wave simulation, our work underscores the boundary dependence of skin effect, and provides a concrete prototype design of NHSE implementable by state-of-the-art of topological metamaterial platforms. 
\end{abstract}


%


\section{Introduction}
%
%

%
%
\IEEEPARstart{T}{opological} band theory, a cornerstone of modern condensed matter physics, has successfully explained a series of novel quantum states, such as topological insulators and topological semimetals with its core principle being the classification of phases using topological invariants~\cite{bansil2016, hasan2010, burkov2016, qi2011, chiu2016}. Later, this powerful theoretical framework was extended to Non-Hermitian (NH) systems, giving rise to the vibrant frontier of NH topological matter~\cite{el2018non, gong2018topological, bergholtz2021}. NH physics is ubiquitous in open systems, for gain and loss are inherent and controllable in photonic and acoustic systems~\cite{parto2020non, hu2021non}, which meanwhile is described by NH Hamiltonians to account for non-equilibrium conditions or finite-lifetime quasiparticles in electronic systems~\cite{kozii2024non}. Even more importantly, non-Hermiticity can be ingeniously harnessed to induce nonperturbtive phenomena and applications, such as novel lasers~\cite{bandres2018, zeng2020}, highly sensitive sensors\cite{budich2020non, lau2018fundamental, hodaei2017enhanced}, and topological energy transport~\cite{longhi2017non, yan2024transport}.

Existing research on NH topology was largely focused on line-gap topology~\cite{kawabata2019symmetry}. In such systems, the energy spectrum possesses a line gap in the complex plane, allowing for the generalization of many concepts from Hermitian topology (e.g., Chern numbers). Within this framework, unique NH phenomena like parity-time (PT) symmetry breaking and exceptional points (EPs) have been extensively studied~\cite{el2018non, bergholtz2021}. EPs are degeneracies unique to NH systems where eigenvalues and eigenvectors coalesce simultaneously, leading to a range of remarkable physical effects such as enhanced sensing~\cite{budich2020non, yuan2023non}. Moreover, non-Hermiticity induces a distinct class of topology absent in Hermitian counterparts, i.e., point-gap topology~\cite{bergholtz2021, kawabata2019symmetry}. Its criterion is whether the energy spectrum forms a non-trivial winding number around a reference point in the complex energy plane. The striking consequence of point-gap topology is the non-Hermitian skin effect (NHSE)~\cite{gong2018topological, yao2018edge, borgnia2020non, okuma2020topological, zhang2020correspondence}. This effect refers to the localisation near boundaries of the bulk states which cease to extend under open boundary conditions. The skin mode is fundamentally different from the boundary states in topological insulators, because the NHSE shrinks almost all bulk states to become localized towards boundaries. This directional accumulation of bulk waves opens new avenues for possible applications for directional transport and energy harvesting. 

Although the NHSE in one-dimensional systems has been widely reported and investigated, its manifestation in higher dimensions is more complex and richer~\cite{kawabata2020higher, zhang2021observation, zhang2022universal}. A key advance is that higher-dimensional NHSE exhibits significant geometry-dependence or anisotropy~\cite{zhang2022universal, zhou2023observation}. Note that the geometry-dependent NHSE is not determined by the overall shape of the sample, but by the direction of its open boundaries. Specifically, if one finds out a nonzero winding number along a closed path in the first Brillouin zone (BZ), the complex energy spectrum corresponding to that path will enclose a non-zero area, and henceforth the skin effect will emerge at the boundaries perpendicular to that direction~[28, 29]. This suggests the potential to actively control wave localization by judiciously designing the lattice boundaries for unprecedented manipulation of waves.
Presently, systematic studies that integrate anisotropic NHSE into concrete classical wave (e.g., photonic) platforms remain rare. In particular, how non-Hermiticity influences point-gap topology in a Lieb lattice, which possesses unique band structures~\cite{guzman2014, mukherjee2015}, and how to realize the anisotropic NHSE, is a pressing question to answer. Inspiring from previous work on the multiple hierarchy of higher-order topology for the Lieb photonic lattice~\cite{xiong2026}, we expect to further investigate its NH properties. Over the simple square lattice, the Lieb lattice structure allows straightforward construction of multiple supercell geometries (square, parallelogram, triangular) within the same unit cell, which is convenient for systematically adjusting the boundary orientation under identical NH settings. NH Lieb lattices have been found to host rich EP topology such as astroid-shaped exceptional loops~\cite{xiao2020exceptional}. Moreover, the geometry-dependent NHSE has recently been experimentally demonstrated in reciprocal two-dimensional lattices~\cite{wang2023experimental}. Differently, our work provides a systematic investigation of geometry-dependent NHSE in a two-dimensional NH Lieb lattice, where the anisotropic skin effect in a photonic crystal (PhC) can be realised by introducing inter-cell NH hopping. Our finding is the prediction and validation of a strongly anisotropic NHSE in a Lieb PhC, which occurs only for specific tilted boundaries, via calculations of the winding number and wavefunction localization in the designed finite-sized structures. This study underscores the critical role of boundary design in controlling wave localization within NH systems. 

\section{Tight binding model (TBM)}\label{Theoretical}
\begin{figure}[hbtp!]
\includegraphics[width=0.48\textwidth]{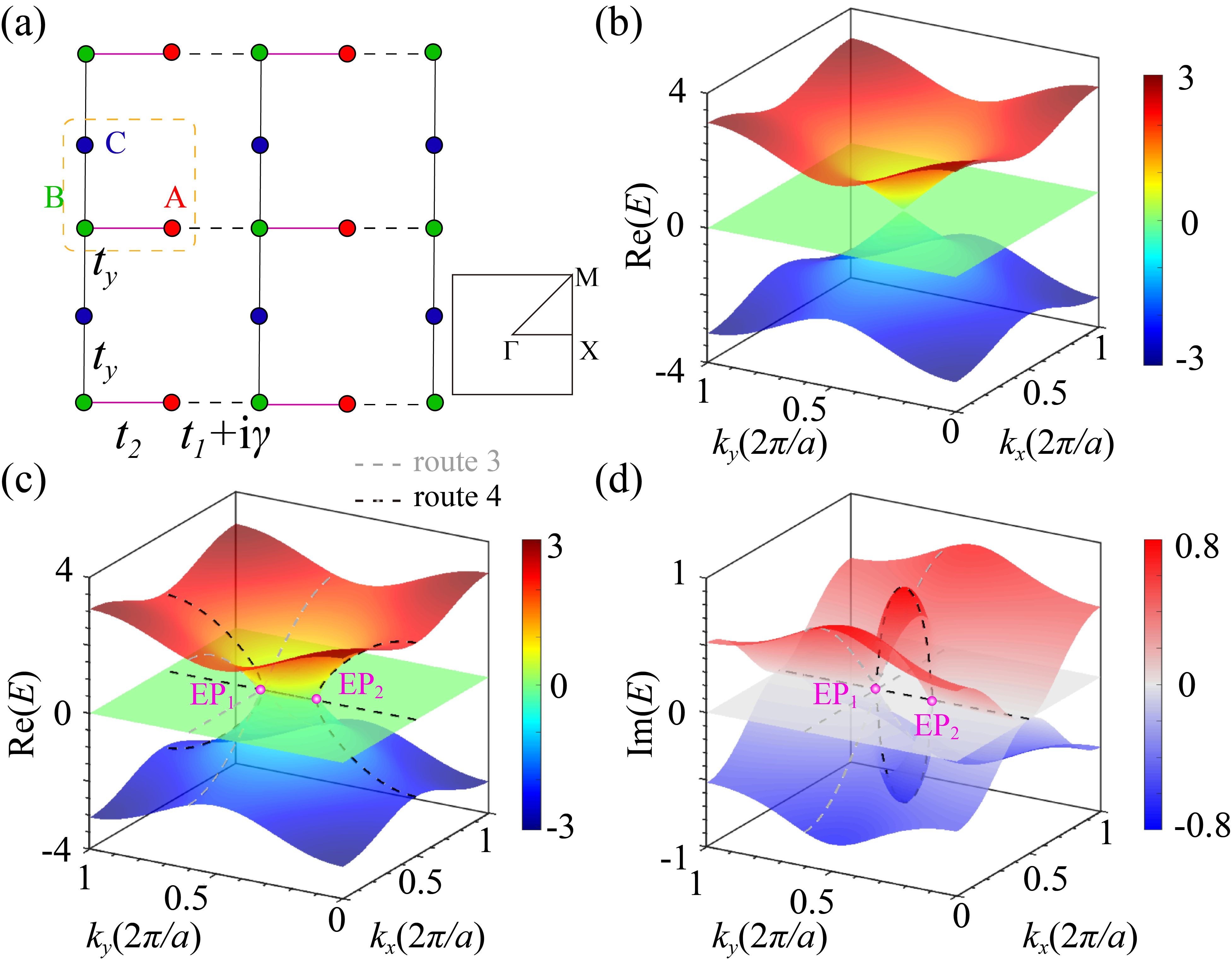}
\caption{\label{fig:fig1}(a) Schematic for the Lieb lattice, where A, B and C represent three sublattices. The unit cell is marked by an orange dashed line, and the inset represents the first BZ of such a unit cell. (b) Band structure of the Hermitian system with $\gamma=0$, the energy is purely real. (c)-(d) Real and imaginary band structure of the NH system with $\gamma=0.8$. The dashed lines represent the energy spectra corresponding to different routes in the first BZ [cf. Fig.~\ref{fig:figA1}(a) in Append.~\ref{EP}]. Parameters: $t_1=t_2=1$, $t_y=1.2$.
}
\end{figure}

Here, we study a pristine Lieb lattice as shown in Fig.~\ref{fig:fig1}(a), whose unit cell includes three sublattices A, B, and C. In this model, we only consider the nearest neighbor hopping. The intra-cell hopping $t_2$ along the horizontal direction is Hermitian, while the inter-cell hopping $t_1+i\gamma$ is NH. The hopping along the vertical direction are reciprocal, with both intra-cell and inter-cell hopping denoted as $t_y$. With the real parameters $t_1, t_2, t_y$ and $\gamma$, the TBM Hamiltonian of our PhC can be written as
\begin{eqnarray}\label{H(k)}
H(\textbf{k})=\left[\begin{array}{ccc}
0&t_2+(t_1+i\gamma)e^{ik_xa}&0\\
t_2+(t_1+i\gamma)e^{-ik_xa}&0&t_y+t_ye^{-ik_ya}\\
0&t_y+t_ye^{ik_ya}&0\end{array}\right], 
\end{eqnarray}
where $a$ is the lattice constant. 
The eigenvalues of Eq.~\eqref{H(k)} are:
\begin{eqnarray}
E_0(\textbf{k})=0,\quad
E_{\pm}(\textbf{k})=\pm\sqrt{M+iN},\label{Ek}
\end{eqnarray}
where
\begin{eqnarray}
M&=&t^2_1+t^2_2-\gamma^2+2t_1t_2\cos k_x+2t^2_y(1+\cos k_y),\nonumber\\
N&=&2\gamma(t_1+t_2\cos k_x). 
\end{eqnarray}
When $\gamma=0$, the system is Hermitian, and the band structure in Fig.~\ref{fig:fig1}(b) shows a typical flat band. When $\gamma\neq0$, the system is NH due to the complex hopping $t_1+i\gamma$, where the $\gamma$ represents gain (or loss) of such a NH system. For the NH system, the band structures are shown in Figs.~\ref{fig:fig1}(c) and (d), there exist non-zero imaginary components in their energy spectrum. More importantly, EPs occur in reciprocal space, where two non-zero eigenvalues and their corresponding eigenstates simultaneously become degenerate [cf. Fig.~\ref{fig:figA1} in Append.~\ref{EP}]. According to Eq.~\eqref{Ek}, the precise parameter conditions for the existence of EPs can be derived, 
\begin{eqnarray}
k_x&=&\pm\arccos\Big(-\frac{t_1}{t_2}\Big),\\
k_y&=&\pm2\arccos \sqrt{\frac{\gamma^2-t^2_2+t^2_1}{4t_y^2}},\nonumber\\
\end{eqnarray}
where the parameter requires $|t_1/t_2\leq 1|$ and $(\gamma^2-t^2_2+t^2_1)/4t^2_y\leq1$ for real values at EPs.

\begin{figure*}[hbtp!]
\includegraphics[width=0.96\textwidth]{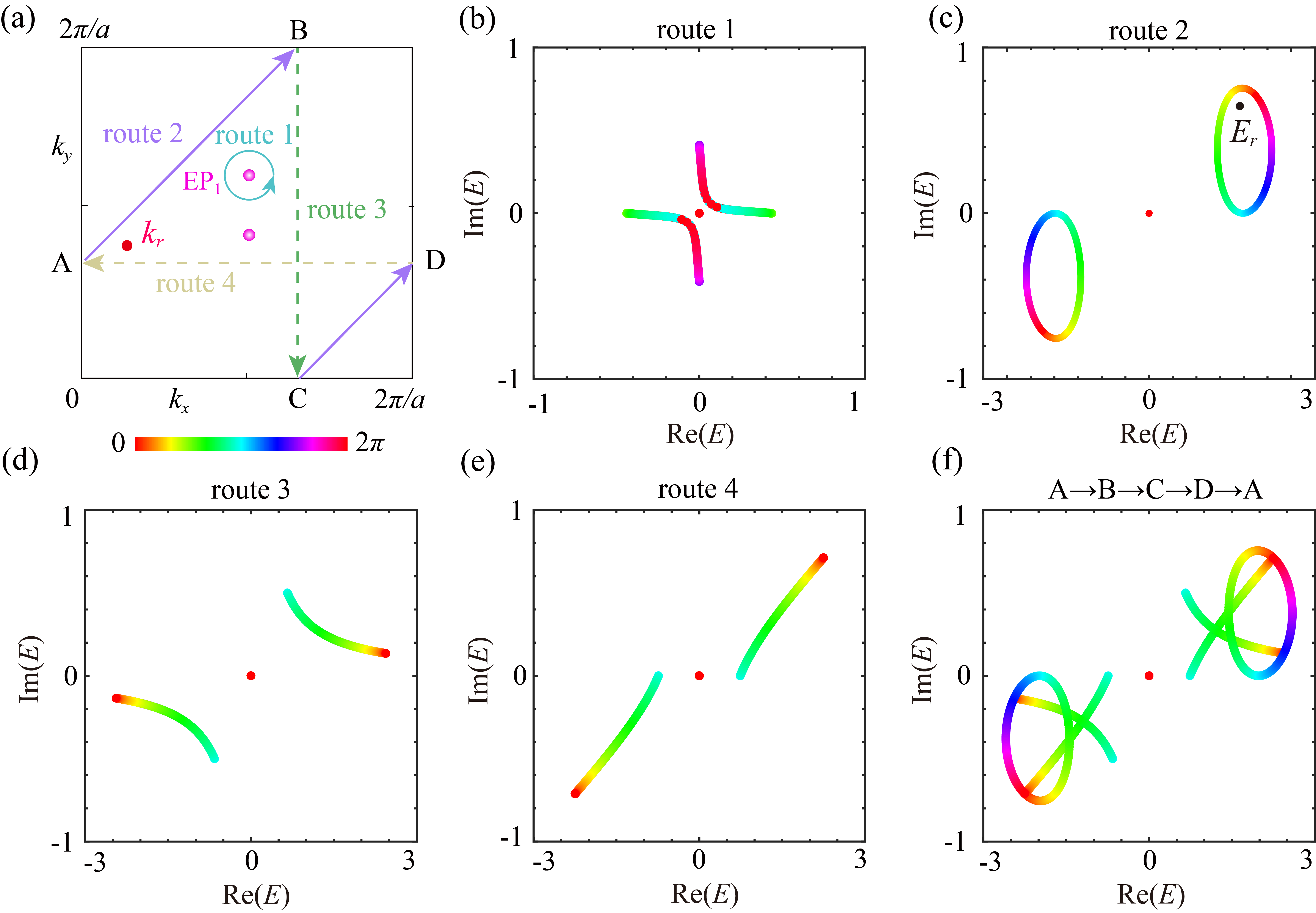}
\caption{\label{fig:fig2}(a) Different routes in the first BZ, the magenta points represent
the EPs, and the red point is the reference point we chosed with
$k_r(0.35\pi/a, 0.8\pi/a)$ and its energy $E_r=1.8+0.65i$. (b)-(e) Complex energy spectra of route 1 to route 4, and (f) corresponds the complex energy spectrum of the total closed route $A\rightarrow B\rightarrow C\rightarrow D\rightarrow A$. All the parameters are same as those in Fig.~\ref{fig:fig1}.
}
\end{figure*}

In general, the system can host up to four EPs. Since we choose the parameters such that $t_1=t_2$, the two solutions for $k_x=\pm\pi$ coincide which results in the system containing only two EPs ($\textrm{EP}_1$ and $\textrm{EP}_2$), as shown in Figs.~\ref{fig:fig1}(c) and (d). These EPs represent topological defects in the NH band structure, typically carrying non-zero topological charges. Here, we calculate the topological charge of EPs~\cite{zhang2022universal, zhou2023observation} using, 
\begin{eqnarray}\label{v(EP)}
\nu(\textbf{k}_{\rm{EP}})=\frac{1}{2\pi i}\oint_{\Gamma_{\rm{EP}}}d\textbf{k}\cdot\nabla_{\textbf{k}}\ln\det[H(\textbf{k})-E(\textbf{k}_{\rm{EP}})],
\end{eqnarray}
where $\Gamma_{\rm{EP}}$ is a counterclockwise closed path that encloses while not passing through the EP, and $E(\textbf{k}_{\rm{EP}})$ is the energy of EP. This topological charge is also the spectral winding number of the closed route. Generally, the EPs profoundly influence the bulk-boundary correspondence and the emergence of the NHSE, thereby providing a crucial theoretical foundation for controlling wave propagation and transport properties. However, in our model, the EPs show trivial topological charge, i.e., $\nu(\textbf{k}_{\rm{EP_{1,2}}})=0$, the route 1 is a small loop around the $\textrm{EP}_1$ to evaluate its topological charge, corresponding the $\Gamma_{\rm{EP}}$ in Eq.~\eqref{v(EP)}, as shown in Fig.~\ref{fig:fig2}(a). For trivial topological charge, the complex energy spectrum of such route 1 shows two open arcs and one point, as shown in Fig.~\ref{fig:fig2}(b), corresponding to the three bands of Lieb lattice. Due to the zero energy flat band, the energy spectrum will show a point in complex space. 

Although the topological charges associated with the EPs are trivial, we find that there exists a reference point $k_r$ in the first BZ which carries a nonzero winding number. This implies that within the BZ, there must exist a closed straight line route $k_n$ exhibiting a nonzero winding number, which describes the spectral winding number around the $E_r$ on the complex plane, which is known as the point gap. Further, there will be a geometry dependence on the NHSE emerges at the open boundaries vertical to the $k_n$ direction~\cite{zhang2022universal, zhou2023observation}. The winding number is defined as
\begin{eqnarray}
\nu(k_r)=\frac{1}{2\pi i}\oint_{L}d\textbf{k}\cdot\nabla_{\textbf{k}}\ln\det[H(\textbf{k})-E(r)],\label{v(kr)}
\end{eqnarray}
where $L$ is the closed straight line route along the $k_n$ direction in the first BZ, $E(r)$ is the energy of reference point $k_r$. Notably, equations~\eqref{v(EP)} and \eqref{v(kr)} show the same mathematical formulation, and only differ in the selection of the integration contour and reference energy. Equation~\eqref{v(EP)} is intendedly to be evaluated for the local topological winding number around EPs, while equation~\eqref{v(kr)} characterizes the global point-gap topology of the system.

\begin{figure*}[hbtp!]
\includegraphics[width=0.96\textwidth]{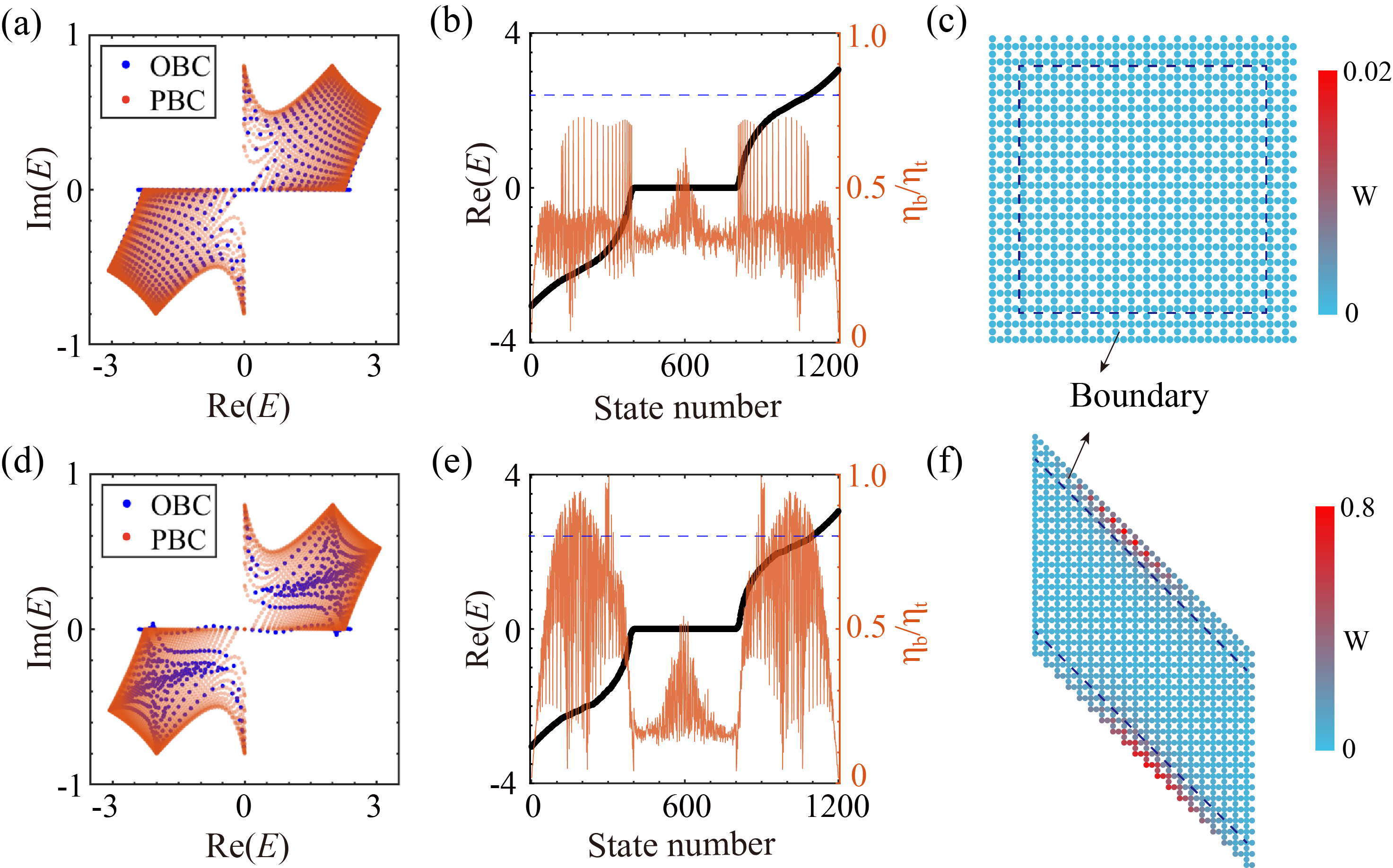}
\caption{\label{fig:fig3}(a) and (d) Spectral area under square and parallelogram shaped structures under OBC. The red points represent the spectral area under periodic boundary condition (PBC). (b) and (e) The real energy spectra of two structures sorted by the real parts of eigenvalues. The orange line represents the ratio of the probability density of the wave function at the boundary to that over the entire structures. The blue dashed line represents the degree of localization $\eta_b/\eta_t=80\%$ of the probability density at the boundary for each eigenstate. (c) and (f) Spatial distribution of all eigenstates $W(x)$ distributions of two shaped structures, both consist of $20\times20$ unit cells. The boundary is defined as having a thickness of two unit cells.
}
\end{figure*}

Here, we select a closed route $A\to B\to C\to D\to A$ in the first BZ, which can be divided into three closed straight line routes 2, 3, and 4, as shown in Fig.~\ref{fig:fig2}(a). The reference point we select is $k_r(0.35\pi/a, 0.8\pi/a)$ and its energy $E_r=1.8+0.65i$ are marked in figure. It should be clarified that a point can be selected as the reference point only if it lies in the first BZ, and its corresponding energy falls inside the complex spectral loop shown in Fig.~\ref{fig:fig2}(c) along with a non‑zero winding number. For route 2, the complex energy spectrum forms a closed loop, resulting in non-zero spectral winding number for the energy of reference point $E_r$, i.e., $\nu(k_r)=1$, as shown in Fig.~\ref{fig:fig2}(c). Differently, the $\nu(k_r)=0$ for both routes 3 and 4, corresponds to no spectral loops at all, as shown in Figs.~\ref{fig:fig2}(d) and (e). And for the closed path shown in Fig.~\ref{fig:fig2}(f), it is used to cross-check and confirm the essential nonzero spectral winding carried route 2 in panel (c).

We now take two shapes of ribbons to test their NHSE: square and parallelogram, respectively. For trivial winding number (e.g. $\nu(k_r)=0$), no skin effect occurs as in the ribbons under rectangle stripe geometry [cf. Fig.~\ref{fig:figA2} in Append.~\ref{ribbons}]. The complex energy spectrum of such square-shaped lattice under open boundary condition (OBC) is shown in Fig.~\ref{fig:fig3}(a).  It is noted that a non-zero area of the complex spectrum under PBC indicates the presence of a geometry-dependent skin effect in the system. A non-zero area enclosed by the complex energy spectrum under PBCs is equivalent to a non-trivial winding number of the spectrum around a reference point inside that area. In NH systems, this non-trivial winding number is the necessary and sufficient condition for the NHSE under open boundaries. Consequently, when the PBC spectrum encloses a finite area, the winding number is guaranteed to be non-zero, and therefore the system exhibit the skin effect. In our model, we can see that the spectral area of the square-shaped structure under PBC is the same as that under OBC, and thus there is no skin effect, as shown in Figs.~\ref{fig:fig3}(b) and \ref{fig:fig3}(c). Figure~\ref{fig:fig3}(b) shows the real spectrum of the square-shaped structure with $20\times20$ units, where $\eta_b$ and $\eta_t$ represent the probability density of the wave function at the boundary and over the entire structure, respectively. Here, the threshold criterion $\eta_b/\eta_t=80\%$ serves as a conventional criterion in the NHSE literature to quantitatively characterize strongly localized skin modes~\cite{zhou2023observation}. And the boundary is defined as having a thickness of two unit cells (e.g. $n=2$). To visualize the skin effect, we calculate the spatial distribution of total eigenstates $W(\textbf{x})$, which is defined as~\cite{zhang2022universal, zhou2023observation}
 \begin{eqnarray}
 W(\textbf{x})=\frac{1}{N}\sum_n|\psi_n(\textbf{x})|^2,
 \end{eqnarray}
 where $\psi_n(\textbf{x})$ is the n-th normalized right eigenstate at site $\textbf{x}$, and N is the number of eigenstates. 

 On the contrary, non-zero spectral winding number indicates the existence of skin effect at the boundaries perpendicular to the route 2 direction [cf. Fig.~\ref{fig:figA3} in Append.~\ref{ribbons}]. We find that a parallelogram-shaped super cell happens to be the case, which includs two inclined boundaries perpendicular to route 2 direction and two vertical boundaries. The spectral area of such a parallelogram-shaped lattice under OBC is shown in Fig.~\ref{fig:fig3}(d). It exhibits a distribution markedly different from that under PBC, and the eigenstates are localized at the two inclined boundaries, consistent with our theoretical expectation. Its skin effect, is manifested in Figs.~\ref{fig:fig3}(e) and \ref{fig:fig3}(f). In particular, there are 146 states with $\eta_b/\eta_t\geq80\%$ are considered skin modes, and the number of skin modes $N_s$ increases linearly as the supercell size V increases, e.g., it satisfies volume law with $\delta N_s\approx0.4896\delta V$ [cf. Fig.~\ref{fig:figA7}(a) in Append.~\ref{robustness}]. Note that the threshold and boundary thickness are not fixed. When we change the threshold or the boundary thickness in a reasonable range, such as $75\%-90\%$ and $n=2,3$, the volume law also satisfies [cf. Fig.~\ref{fig:figA5} in Append.~\ref{robustness}]. For comparison, in the square structure, although there exist larger boundary localized states, they are not skin modes but localized states, because they fail the volume law [cf. Fig.~\ref{fig:figA7}(b) in Append.~\ref{robustness}].

\section{NHSE in Lieb PhC}
In Sec.~\ref{Theoretical}, we start from a TBM and demonstrate that introducing NH hopping between unit cells of a Lieb lattice can give rise to a geometry-dependent skin effect. To realize this theory in PhC, we map the TBM model in Sec. II to a concrete PhC design structure. The key for this realisation lies in introducing NH coupling in PhCs. Here, we introduce the imaginary part of the refractive index to induce attenuation during propagation in the medium, thereby creating NH coupling. 

\begin{figure}[hbtp!]
\includegraphics[width=0.48\textwidth]{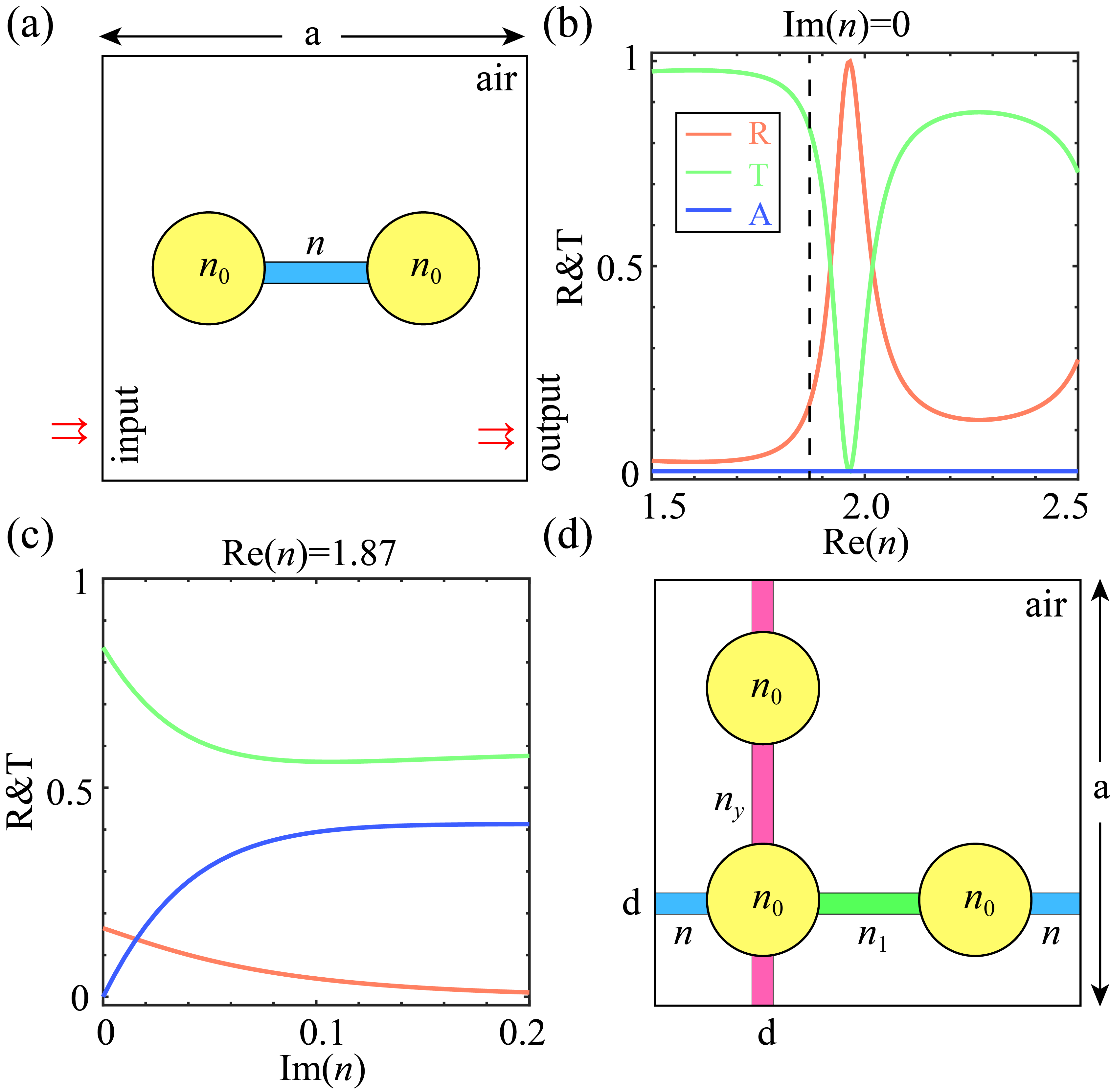}
\caption{\label{fig:fig4}(a) A model for verifying the impact of refractive index on the energy propagation of electromagnetic waves. (b) The distributions of the reflection (R), transmission (T), and absorption (A) rates of the wave field with respect to variations in the refractive index $\rm{Re}(n)$ and (c) $\rm{Im}(n)$ of the rectangular dielectric media. (d) A PhC model based on the Lieb lattice, which is analogous to the TBM. Parameters: $n_0=3.28$, $n=1.87+0.1i$, $n_y=2$, and $d=0.05a$.}
\end{figure}

For a plane electromagnetic wave in transverse magnetic (TM) mode, its electric field can be expressed as
\begin{eqnarray}
\textbf{E}(\textbf{r},t)=E_0e^{i\textbf{k}\cdot\textbf{r}-i\omega t}\hat{x},\label{E}
\end{eqnarray}
For simplicity, we set $\textbf{k}\cdot\textbf{r}=kz$, the wave number $k=n\omega/c$, where n is the refractive index of the medium, c is the speed of the wave in vacuum. When the $n$ is a real number, there is no gain or loss when the electromagnetic wave passes through the media. When the $n$ is a complex number, i.e., $n=n_0+in_1$, with $n_{0,1}>0$. Then equation~\eqref{E} can be written as
\begin{eqnarray}
\textbf{E}(\textbf{r},t)=E_0e^{i(n_0\omega z/c-\omega t)}e^{-n_1\omega z/c}\hat{x}.\label{E1}
\end{eqnarray}
It means that when an electromagnetic wave propagates through a medium with a complex refractive index, its amplitude undergoes exponential decay, then corresponding decay coefficient is $n_1\omega z/c$. To verify this, we design a simple unit cell for our PhC model, as shown in Fig.~\ref{fig:fig4}(a). It consists of two circular dielectric rods with radii $r=0.13a$ and refractive index $n_0=3.28$. The two dielectric rods (GaAs) are connected by a rectangular dielectric media serving as a coupler. The background is set to air. Now, we set the left boundary as the input port and the right boundary as the output port. When a plane wave is input from the left boundary and output from the right boundary, the distributions of the reflection (R), transmission (T), and absorption (A) rates of the wave field with respect to variations in the refractive index $n$ of the rectangular dielectric are shown in Fig.~\ref{fig:fig4}(b). Here, we first assume that the imaginary part of the refractive index $\rm{Im}(n)=0$, and the absorption rate is always zero, which implies that no energy loss occurs. Here we select an optical glass with refractive index $\rm{Re}(n)=n_0=1.87$ marked in dashed line in Fig. 4(b). Next we introduce imaginary refractive index $\rm{Im}(n)=0.1$ for the reason below. As the $\rm{Im}(n)$ gradually increases, the absorption rate also rises, indicating a corresponding increase in energy loss until it eventually stabilizes at $\rm{Im} (n)\sim 0.1$, as shown in Fig.~\ref{fig:fig4}(c). Based on this model, we analogously designed a Lieb PhC structure with NH intercell coupling, as shown in Fig.~\ref{fig:fig4}(d). For the mapped parameters, we assume that the refractive index $n_i$ and the hopping coefficient $t_i$ follow an approximate linear relation ($n_i\propto t_i$, where $t=1,2,y$), based on which the refractive index parameters are determined.

\begin{figure*}[hbtp!]
\includegraphics[width=0.94\textwidth]{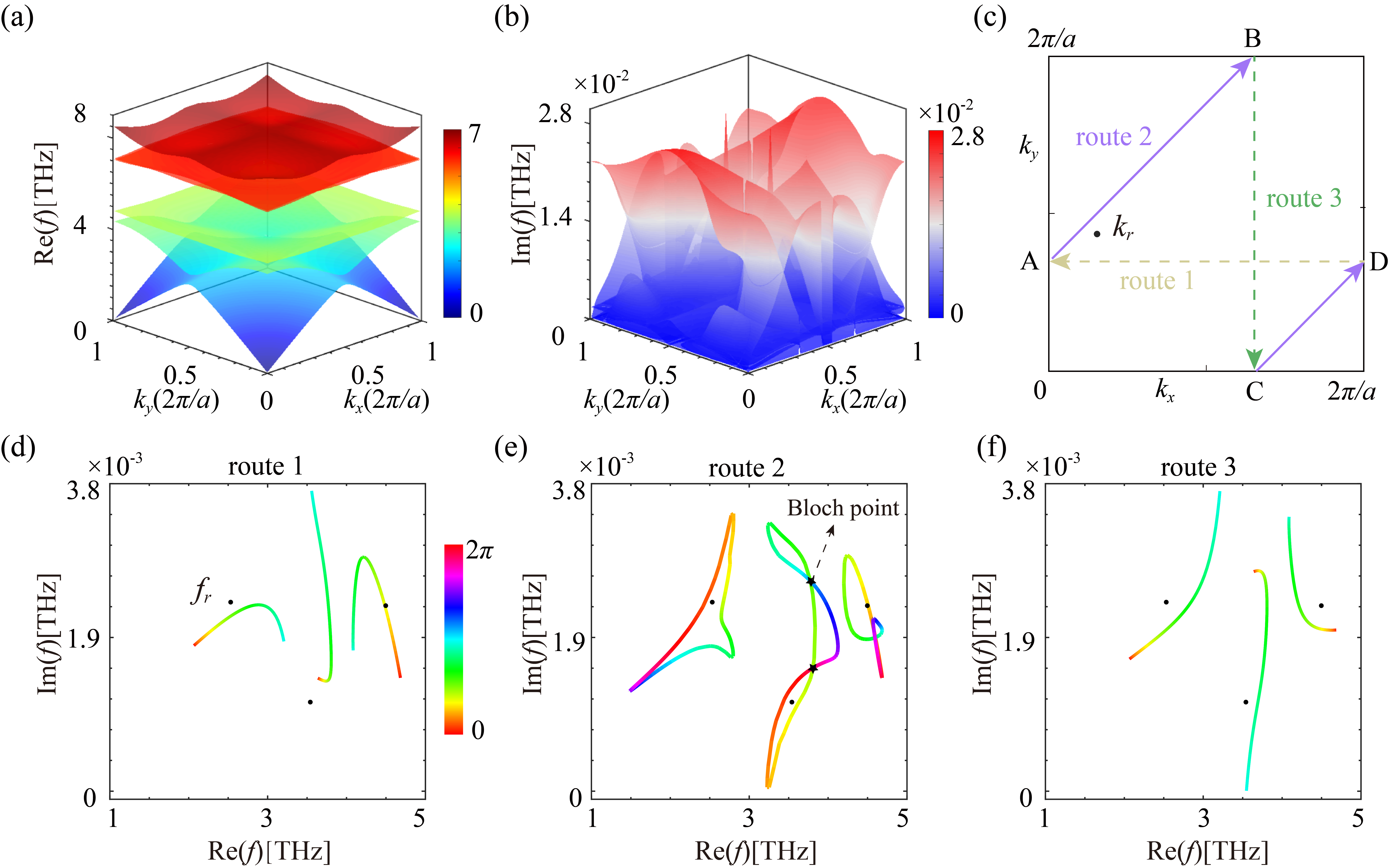}
\caption{\label{fig:fig5}(a)-(b) The real and imaginary parts of the band structures for the Lieb PhC shown in Fig.~\ref{fig:fig4}(d). (c) The first BZ of the Lieb PhC, and we select three closed straight line routes, corresponding to the complex spectra of the three bands are shown in (d)-(f). The reference point we chose is $k_r(0.35\pi/a,0.8\pi/a)$, the complex frequences $f_r$ of the $k_r$ marked by black dots, and the self-intersection Bloch points are marked as black pentagram.
}
\end{figure*}

For this Lieb PhC, the band structures are shown in Figs.~\ref{fig:fig5}(a) and~\ref{fig:fig5}(b). We calculated five energy bands, among which the first three bands overlap in the complex frequency plane and exhibit no clean real-frequency gap. Unlike the TBM, there exist no EPs. The first BZ is shown in Fig.~\ref{fig:fig5}(c), where we select the same three closed path as shown in Fig.~\ref{fig:fig2}(a) and the corresponding complex spectra are shown in Figs.~\ref{fig:fig5}(d)-(f) [routes 1-3] respectively. The reference point is still $k_r(0.35\pi/a,0.8\pi/a)$, with the complex frequencies of $k_r$ for first three bands are marked in figure. Analogous to Eq.~\eqref{v(kr)}, the spectral winding number with respect to $k_r$ in photonic system can be defined as~\cite{zhong2021nontrivial}
\begin{eqnarray}
\nu(k_r)=\oint_L\frac{d\textbf{k}}{2\pi}\nabla_{\textbf{k}}\arg[f_n(\textbf{k})-f_r], \label{eqR1}
\end{eqnarray}
where $L$ is still the closed straight route in the first BZ, $f_n(\textbf{k})$ is the eigenfrequency corresponding to the $n$-th band of PhC, and $f_r$ is the eigenfrequency at the reference point $k_r$.

For simplicity, we focus on the first three photonic bands. They manifest clear point-gap topology and nontrivial spectral winding number, which are sufficient to cater for NHSE. For route 1, the complex spectrum forms an open curve with a winding number of zero, as shown in Fig.~\ref{fig:fig5}(d). This implies that when open boundaries are applied perpendicular to route 1, the skin effect will not show along those boundaries. Particularly, for route 2, the complex spectrum of the first three bands winds into a closed loop, with nonzero spectral winding number. The nonzero spectral winding number implies that introducing boundaries perpendicular to route 2 will result in occurring of the skin effect. According to equation~\eqref{eqR1}, the complex spectrum of the first band forms a closed loop with $\nu=-1$. Interestingly, the complex spectra of the second and third show more complicated winding patterns: the second band forms a closed loop which encloses three separate regions (holes) with the winding number with respect to $k_r$ being $\nu=1$, as shown in Fig.~\ref{fig:fig5}(e). Moreover, the band features two self-intersections at the Bloch points~\cite{song2019non, longhi2019probing, zhong2021nontrivial}. The two complex frequency spectrum regions adjacent to the Bloch point are associated with opposite winding directions, and this behavior leads to the interpretation of the Bloch point as indicative of a bipolar skin effect~\cite{song2019non}. We note that Bloch points have recently been observed experimentally in topological acoustic systems governed by TBMs~\cite{zhang2021observation, zhou2023observation, zhang2021acoustic}, and our work demonstrates their feasibility in reciprocal PhC structures too. As for route 3, the complex spectrum still forms an open curve with trivial winding, as shown in Fig.~\ref{fig:fig5}(f), so there is no skin effect emerge along the boundaries which are perpendicular to route 3 directions. 

\begin{figure*}[hbtp!]
\includegraphics[width=0.94\textwidth]{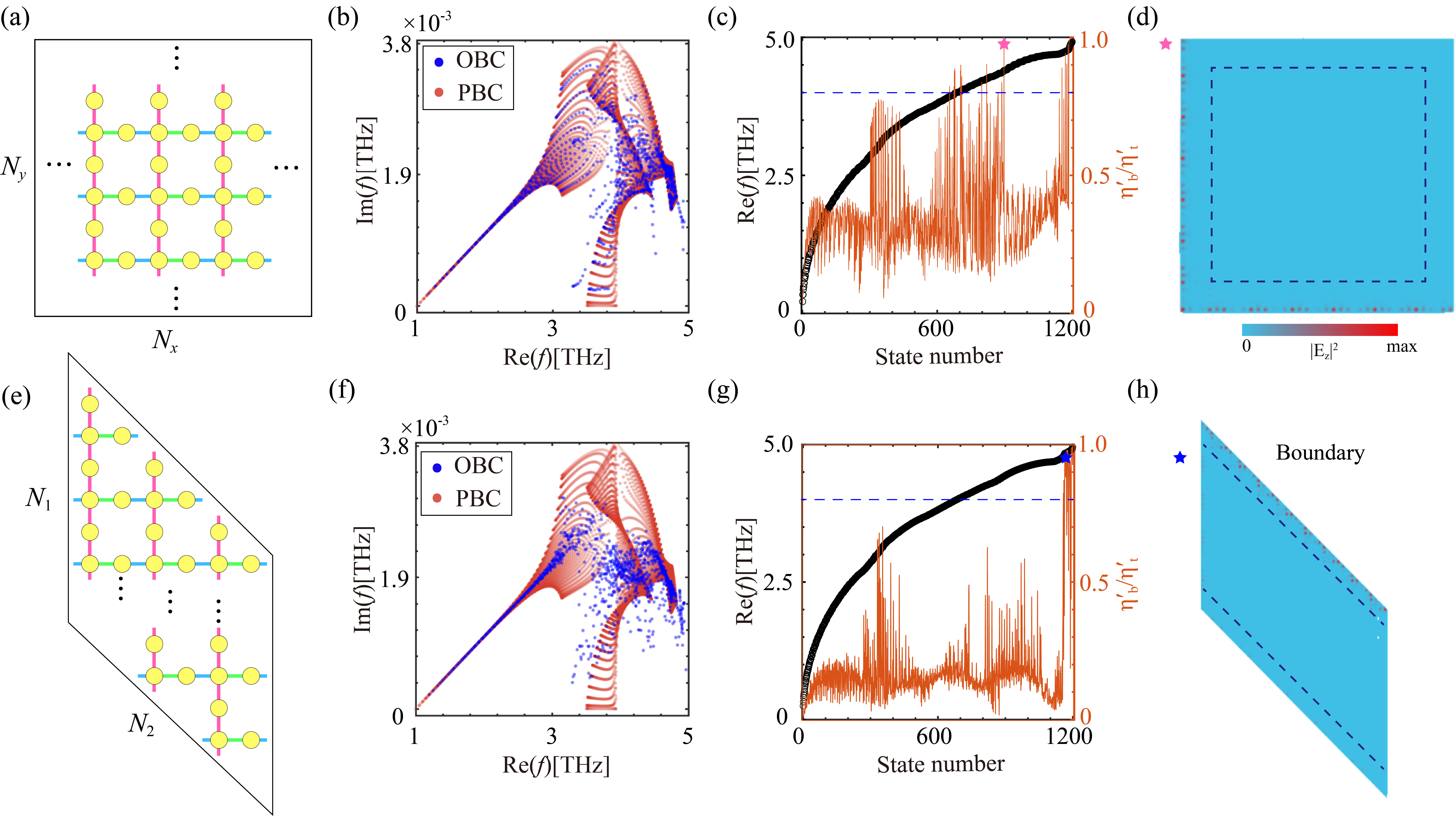}
\caption{\label{fig:fig6}(a) and (e) Two types of super cell with different open boundaries. (b) and (f) PBC and OBC complex spectra of two supercells. (c) and (g) The OBC spectra of two types of super cell, where $\eta'_b$ and $\eta'_t$ represent the electric field energy density at the boundary and of the entire supercell, respectively. The corresponding state numbers are 886 and 1188. (d) and (h) The field distributions of states 1188 and 1170, marked by magenta and blue pentagram, respectively. The number of unit cells $N_i=20$, and the boundary is defined as having a thickness of two unit cells.} 
\end{figure*}

To verify the PhC skin effect, we have designed two supercells with different open boundaries, as shown in Fig.~\ref{fig:fig6}(a) and \ref{fig:fig6}(e). Both structures are composed of $20\times20$ unit cells. For the square-shaped supercell, the complex energy spectrum under OBCs is almost identical to that under PBCs in area, as shown in Fig.~\ref{fig:fig6}(b), and there is no skin effect. Figure~\ref{fig:fig6}(c) shows the real spectrum corresponding to the square supercell. Here, we calculated 1200 eigenfrequencies and computed the electric field  for each eigenfrequency, where $\eta'_b$ represents the electric field energy at the boundary, and $\eta'_t$ represents the electric field energy of the entire supercell. The ratio of the two indicates the degree of localization of an individual electric field at the boundary. Here, we take a thickness of 2 unit cells as the boundary and choose $80\%$ threshold to define the skin modes. It should be noted that in the square-shaped supercell, there exist states with a ratio $\eta'_b/\eta'_t>80\%$. Although their eigenstates are localized at the boundary, as shown in Fig.~\ref{fig:fig6}(d), these states are not regarded as skin modes but localized states, because the number of these states does not obey the volume law [cf. Fig.~\ref{fig:figA4}(c) in Append.~\ref{Volume}]. For the parallelogram-shaped structure, its complex energy spectrum under OBCs is clearly different from that under PBCs, indicating that the conventional bulk-boundary correspondence breaks down, as shown in Fig.~\ref{fig:fig6}(f). The corresponding real energy spectrum of the parallelogram structure is shown in Fig.~\ref{fig:fig6}(g). Here, we also calculated the degree of localization $\eta'_b/\eta'_t$ of the electric field energy at the boundary for each eigenstate, with the boundary defined as having a thickness of 2 unit cells (e.g. $n=2$). There are 33 states that satisfy the condition $\eta'_b/\eta'_t\geq 80\%$ and are considered as skin modes, which obey the volume law [cf. Fig.~\ref{fig:figA4}(c) in Append.~\ref{Volume}]. We selected the skin mode as shown in Fig.~\ref{fig:fig6}(h), and the electric field is shown to localise at the inclined boundary, but almost no electric field distribution shown at the vertical boundary. This is consistent with our previous analysis of the complex energy spectrum. Moreover, we have tested multiple threshold values within a reasonable range ($75\%-90\%$) and different boundary thickness ($n=2$ and $n=3$), as shown in Fig.~\ref{fig:figA6}. The test results show that the overall geometric-dependent behaviour of the skin effect remains consistent, demonstrating that our main conclusion is robust against moderate variations of the threshold and boundary thickness.

In addition, we plot the spatial distribution of all eigenstates $W(x)$ distributions of two shaped structures [cf. Figs.~\ref{fig:figA4}(a) and \ref{fig:figA4}(b) in Append.~\ref{Volume}]. It should be noted that among the calculated 1200 states, only 33 are skin modes are found. Therefore, the skin effect is scarcely visible in $W(x)$. By the way, our conclusions are robust when examining individual eigenstates, which show that electromagnetic waves are localized along inclined boundaries. The averaged quantity $W(x)$ provides a clear macroscopic overview of spatial localization, which also aligns with inspection of single eigenstates.

\section{Experimental remark}
Our work based on theoretical modeling and numerical simulations, can be experimentally feasible according to our proposed Lieb PhCs. In the unit cell, we may use gallium arsenide (GaAs) for refractive index of $n_0=3.28$, quartz glass for $n_y=2$, and the complex refractive index $n_1=1.87+0.1i$ by depositing a thin lossy layer (such as a chromium oxide film) on optical glass. Thus the entire structure can be fabricated via conventional electron-beam lithography and dry-etching techniques~\cite{mizeikis2003silicon, gan2012designs, xie2019visualization}. Furthermore, numerical verification across $10\times10$ to $30\times 30$ supercells confirms the geometry-dependent NHSE is robust to small structural deviations (e.g., lattice constant, rod radius), and simulations with $\rm{Im}(n)=0.1$ and $\rm{Im}(n)=1$ show stable volume-law scaling of skin modes, indicating tolerance to moderate material loss. It is noted that while the TBM uses discrete complex hopping for NH coupling and the PhC employs uniform material loss as a macroscopic simplification, both reciprocal, the mapping captures key physics (spectral winding, NHSE) without fully replicating the microscopic NH distribution of the ideal model. Therefore the Lieb PhC provides a feasible prototype for future experiments focusing on precise control of material loss and boundary orientation to verify our predictions.

\section{Conclusion}
In summary, we demonstrate a geometry-dependent NHSE in a two-dimensional Lieb PhC. By introducing NH coupling via complex refractive indices, we propose a photonic design of NH hopping. Through spectral winding number analysis, we predicted that the skin effect emerges when boundaries are oriented along specific tilt directions ($y=\pm x$) with nontrivial point-gap topology. Full-wave simulations on parallelogram structures confirmed this behaviour: eigenstates localize at inclined boundaries, and the number of skin modes obey the volume law, distinguishing them from localised boundary states. Our work provides a practical photonic platform for exploring NH topology and geometry-dependent wave localization, with potential applications in directional energy harvesting and sensing.

\appendices
\section{Examination of EPs}\label{EP}
\renewcommand{\thefigure}{A\arabic{figure}} 
\setcounter{figure}{0} 

\begin{figure*}[hbtp!]
\includegraphics[width=0.94\textwidth]{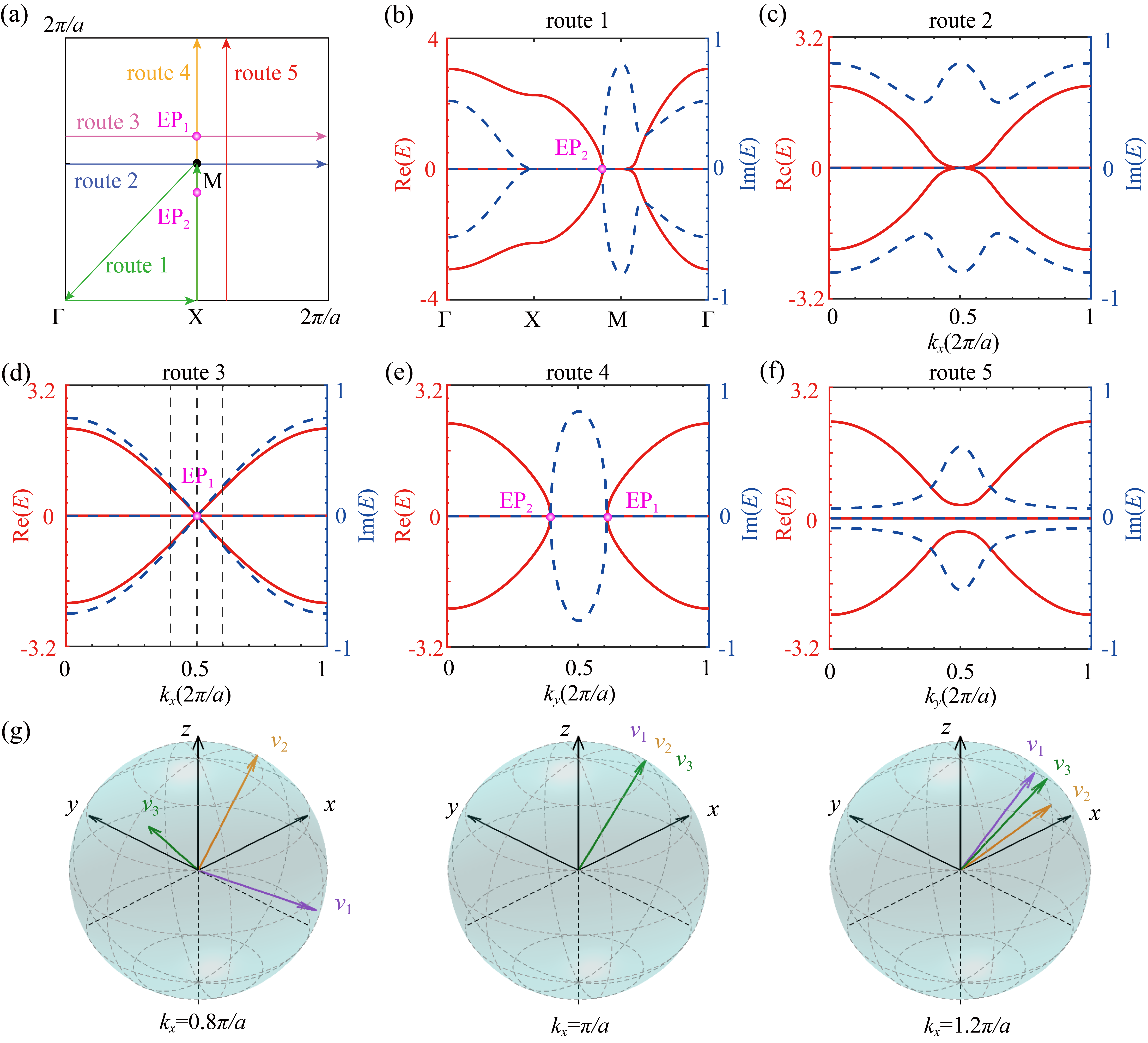}
\caption{\label{fig:figA1}(a) Different routes in the first BZ, the magenta points represent the EPs ($EP_1$ and $EP_2$). (b)-(f) The real and imaginary band structures of different routes in (a). All the parameters are same as those in Fig.~\ref{fig:fig1}. (g) The normalized eigen-vector $v_i(i=1,2,3)$ spatial distributions of the states near the $EP_1$ shown in (d). 
}
\end{figure*}

In this Append., we selected a series of closed routes in the first BZ to verify our calculation of EPs, and the corresponding energy band diagrams for each contour are plotted in Fig.~\ref{fig:figA1}. For route 1, we select high-symmetry points ($\Gamma$, X, and M) in k-space to form a closed loop, where the segment from X to M passes through $\textrm{EP}_2$. The corresponding real and imaginary parts of the energy bands will exhibit degeneracy at $\textrm{EP}_2$, as shown in Fig.~\ref{fig:figA1}(b). For route 2, we fix $k_y=\pi/a$ and define $k_x$ from 0 to $2\pi/a$. This route will pass through high-symmetry point M, and the real part of energy bands exhibit degeneracy, but not the imaginary part of energy bands exhibit, as shown in Fig.~\ref{fig:figA1}(c). So the point M is not an EP. For routes 3 and 4, they pass through one EP and two EPs respectively, corresponding to one and two degeneracy points in band, as shown in Figs.~\ref{fig:figA1}(d)-(e). As for route 5, it does not pass through any high-symmetry points or EPs, and its energy bands exhibit no degeneracy, as shown in Fig.~\ref{fig:figA1}(f). Figure.~\ref{fig:figA1}(g) shows the normalized eigenvector spatial distributions corresponding to the eigenvalues near the $\textrm{EP}_1$. It can be observed that at the $\textrm{EP}_1$, the eigenvectors are also degenerate.

\section{NHSE in the 1D ribbons}\label{ribbons}
In this Append., we investigate the ribbons under different shaped structures to verify the existence of skin effect in 1D ribbon. Specifically, we consider a rectangle and a parallelogram-shaped ribbons.

\begin{figure*}[hbtp!]
\includegraphics[width=0.94\textwidth]{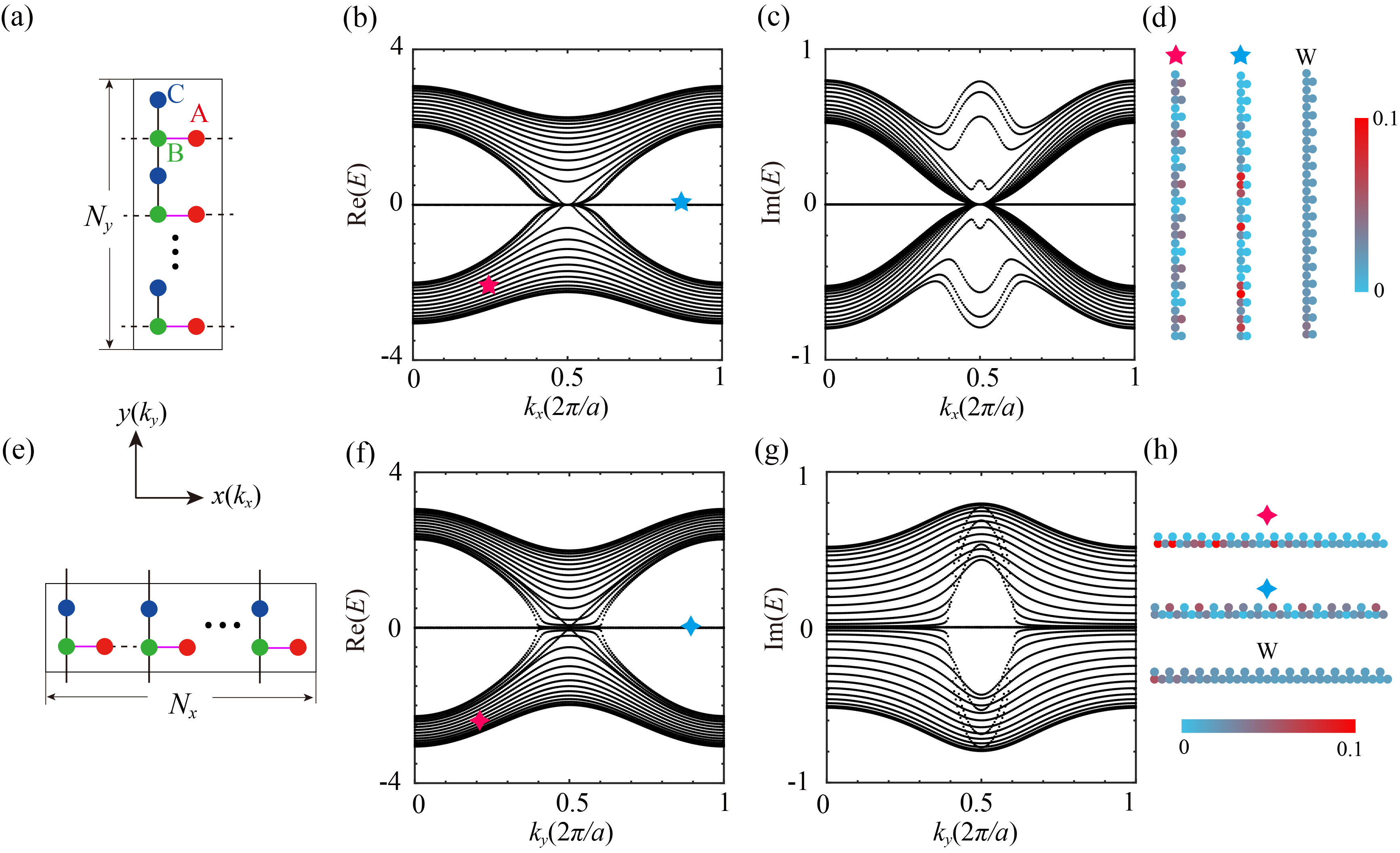}
\caption{\label{fig:figA2}(a) and (e) Super cell structures when we open vertical and horizontal boundaries.
(b)-(c) and (f)-(g) The real and imaginary energy projected bands of two types of super cell. (d) and
(h) The eigenstate distributions of marked and total eigenstates $W(\textbf{x})$ distributions of two ribbons.
}
\end{figure*}

Firstly, for the ribbons under rectangle-shaped, one model is designed with OBCs along the $x$ direction and PBCs in the $y$ direction, as shown in Fig.~\ref{fig:figA2}(a), where $N_y=16$ is the number of unit cell. The projected band of such super cell are shown in Figs.~\ref{fig:figA2}(b) and (c), which show zero-energy flat bands. Here, we present the spatial distribution of eigenstates corresponding to two points, as shown in Fig.~\ref{fig:figA2}(d). Additionally, we determine an eigenstate $\psi(\textbf{x})$ as a skin mode when it satisfies
\begin{eqnarray}
\sum_{x\in\mathcal{B} }|\psi(\textbf{x})|^2\geqslant 80\%,
\end{eqnarray}
where $\mathcal{B}$ represents the boundary region we define. Based on this criterion, the two eigenstates in Fig.~\ref{fig:figA2}(d) are not skin mode but bulk states. Then, we calculate the spatial distribution of total eigenstates $W(\textbf{x})$ for such 1D ribbon, which is defined by the sum of all normalized right eigenstates in the ribbon for all $k_x$. The $W(\textbf{x})$ is almost uniform in the whole ribbon, indicating that there is no skin effect. Similarly, when the ribbon model is designed with OBCs along the $y$ direction and PBCs in the $x$ direction in Fig.~\ref{fig:figA2}(e), the corresponding projected bands in Figs.~\ref{fig:figA2}(f) and (g), manifest no skin effect, as shown in Fig.~\ref{fig:figA2}(h).

\begin{figure*}[hbtp!]
\includegraphics[width=0.94\textwidth]{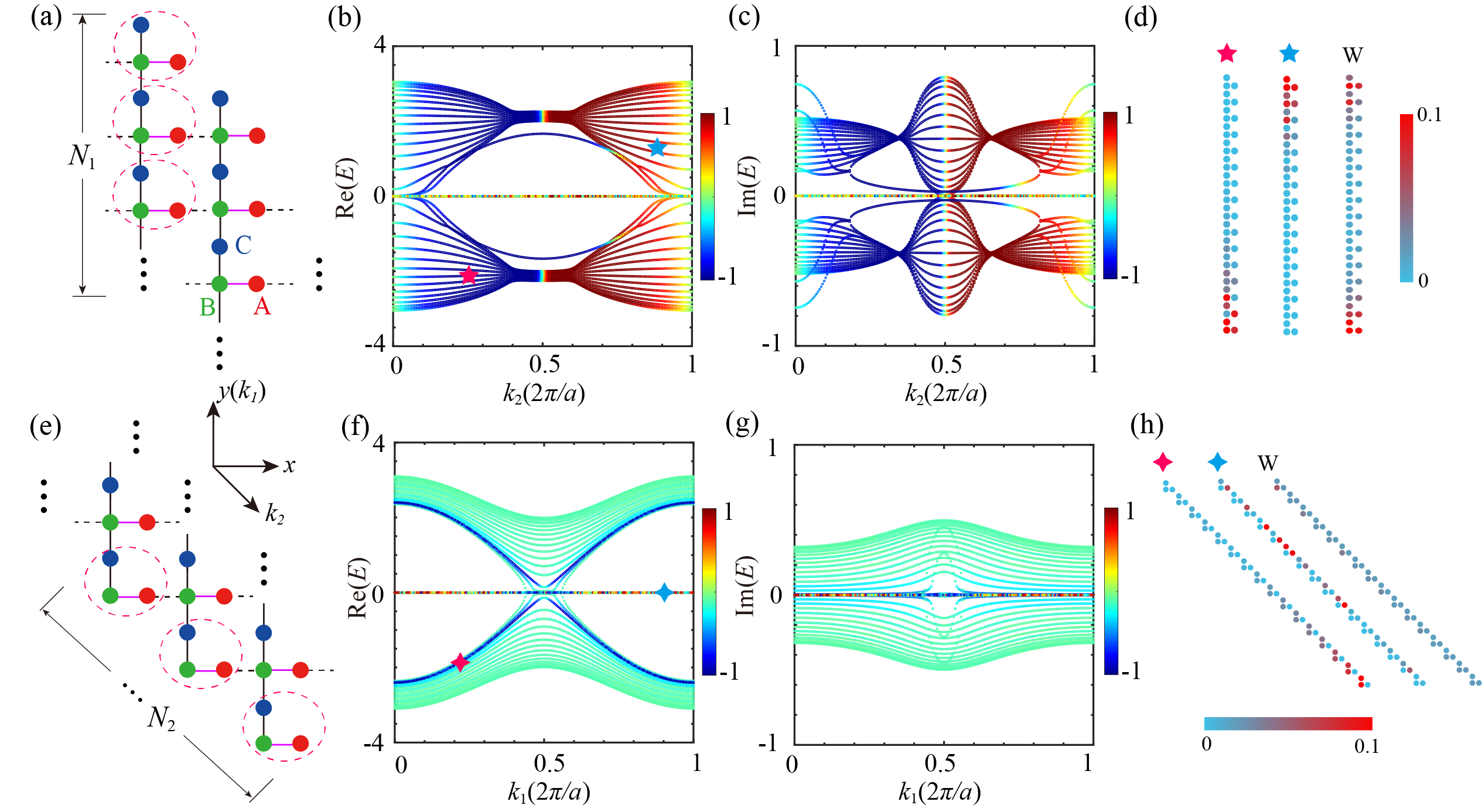}
\caption{\label{fig:figA3}(a) and (e) Super cell structures when we open $k_1$ and $k_2$ boundaries. (b)-(c) and
(f)-(g) The real and imaginary energy projected bands of two types of super cell. (d) and (h) The eigenstate distributions of marked and total eigenstates $W(\textbf{x})$ distributions of two ribbons.
}
\end{figure*}

Secondly, for the parallelogram-shaped ribbons, one designs OBCs along the $k_2$ direction and PBCs along the $y$ (or $k_1$) direction, as shown in Fig.~\ref{fig:figA3}(a) with $N_1=16$. Although the same structure as in Fig.~\ref{fig:figA2}(a), the projected bands exhibit a different distribution due to different PBCs, as shown in Figs.~\ref{fig:figA3}(b) and (c). Interestingly, in this case, the eigenstates of projected bands at almost all k-points are localized at the top and bottom boundaries of the ribbon, manifesting a skin effect. We also select two k-points and plot their spatial distributions of eigenstates, which are indeed localized at the top and bottom boundaries of the ribbon, as shown in Fig.~\ref{fig:figA3}(d). Furthermore, we calculate the spatial distribution of total eigenstates $W(\textbf{x})$, pinpointing the skin effect. Furthermore, another structure is designed with OBCs along the $k_1$ direction and PBCs along the $k_2$ direction, as shown in Fig.~\ref{fig:figA3}(e), where $N_2=16$. We find that there is no skin effect in this case. Although the eigenstates at a few k-points are localized at the boundaries, the overall spatial distribution remains uniform, as shown in Figs.~\ref{fig:figA3}(f)-(h). In such a structure, for the two vertical boundaries, the trivial spectral winding number indicates no skin effect. 

In summary, by opening boundaries along different directions, we observe the presence of boundary-dependent skin effects. Specifically, when boundaries along the horizontal and vertical directions are open, no skin effect emerges, as the spectral winding numbers in both directions are zero. However, when boundaries along the $y=-x$ direction are open, a non-zero spectral winding number leads to the appearance of a skin effect along these boundaries [cf. Fig.~\ref{fig:fig2}].

\section{Volume law}\label{Volume}
In 2D systems exhibiting geometry-dependent skin effects, not all bulk states are skin modes. The number of skin modes satisfies the volume law, meaning that the number of skin modes scales with the volume of the system, i.e.,  
\begin{eqnarray}
\delta N_s\propto\delta V,
\end{eqnarray}
where $N_s$ represents the number of states which satisfies $\eta'_b/\eta'_t>80\%$, and $V=N_x\times N_y$ represents the volume (the total number of unit cells). 

\begin{figure*}[hbtp!]
\includegraphics[width=0.94\textwidth]{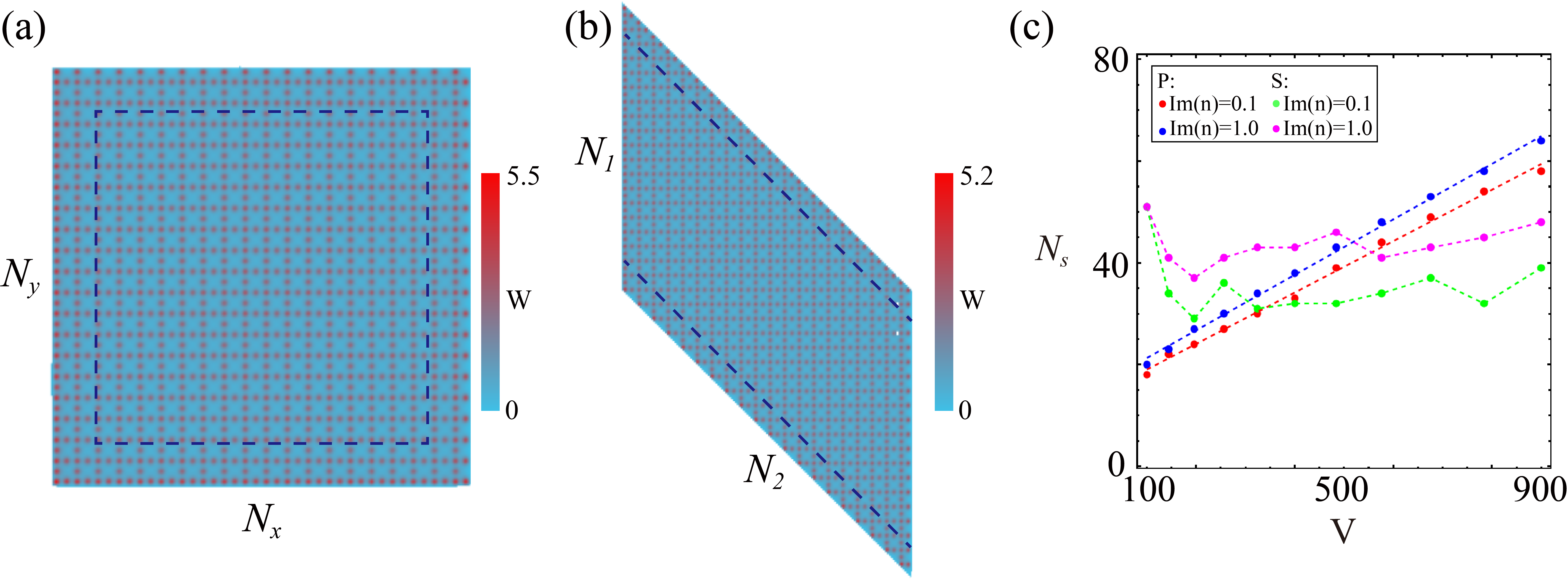}
\caption{\label{fig:figA4}(a)-(b) Spatial distribution of all eigenstates $W(x)$ distributions of two shaped structures in Fig.~\ref{fig:fig6}. (c) The volume law in our system, the red and blue points is plotted by calculating data and the dashed line is the fitting curve showing $\delta N_s\approx0.066\delta V$($\delta N_s\approx0.048\delta V$) for $\rm{Im}(n)=1$($\rm{Im}(n)=0.1$).
}
\end{figure*}

To verify the volume law, we calculate two shaped structures with different volume ($V=10\times10$, $12\times12$...$30\times 30$), and discuss two case with $\rm{Im}(n)=1$ and $\rm{Im}(n)=0.1$, respectively. We obtain the $N_s$ under different geometry size, as shown in Fig.~\ref{fig:figA4}(c). Here, we also take a thickness of 2 unit cells as the boundary thickness. The red and blue dashed line is the fitting curve for these data points with $\rm{Im}(n)=1$ and $\rm{Im}(n)=0.1$, which shows that the $N_s$ of parallelogram-shaped structure satisfies volume law while the square-shaped structure does not. For parallelogram-shaped structure, the relationships between $N_s$ and V satisfy $\delta N_s\approx0.066\delta V$($\delta N_s\approx0.048\delta V$) for $\rm{Im}(n)=1$($\rm{Im}(n)=0.1$). For the square-shaped structure, it is evident that $N_s$ and V break the linear relation. Therefore, we consider that the states corresponding to $N_s$ in the square-shaped structure are not skin modes but rather localized states. 

\section{Robustness verification}\label{robustness}
To further verify the robustness of the volume law followed by the number of skin modes in the geometry-dependent NHSE, we benchmark the influence of different boundary localization thresholds and boundary thickness on the volume law, ensuring that the our main conclusions stands with no regard to moderate perturbations in parameter selection. 

\begin{figure}[hbtp!]
\includegraphics[width=0.48\textwidth]{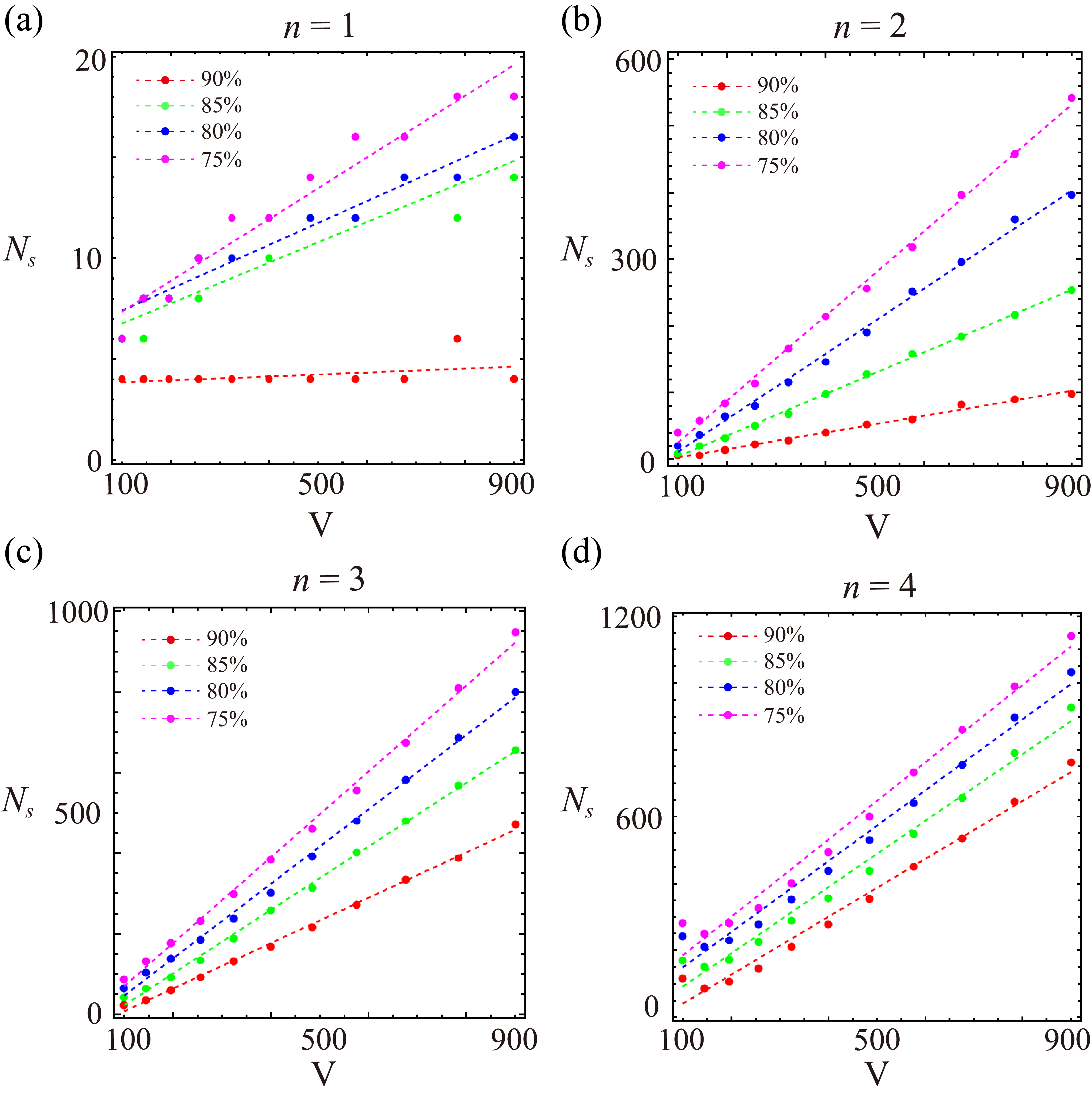}
\caption{\label{fig:figA5}Volume law of parallelogram geometry (for TBM) with different boundary thickness, where $n$ denotes the thickness of the boundary measured in unit cells. The markers correspond to different threshold values adopted for cases $\eta_b/\eta_t$.
}
\end{figure}

In the main text, we adopted a boundary localization threshold (the ratio of boundary probability density/electric field energy density to the total) as $\geq80\%$, and define the boundary thickness as 2 unit (n=2) to identify skin modes. We verify that the number of skin modes \( N_s \) in the parallelogram structure satisfies a linear volume law with the system volume \( V \) (\( V = N_x \times N_y \)). To verify its robustness, we selected different thresholds within a reasonable range $(75\%, 80\%, 85\%, 90\%)$ and boundary thicknesses ($n=1,2,3,4$), recalculated the number of skin modes for supercells of different sizes from $10\times10$ to $30\times30$, and analyzed the linear relationship between \( N_s \) and \( V \), as shown in Fig.~\ref{fig:figA5}. The results indicate that the volume ratio is well kept when the boundary thickness is set to 2 or 3 unit cells. By contrast, the volume ratio breaks down when the boundary thickness is too small ($n=1$) or too large ($n=4$).\\

\begin{figure}[hbtp!]
\includegraphics[width=0.48\textwidth]{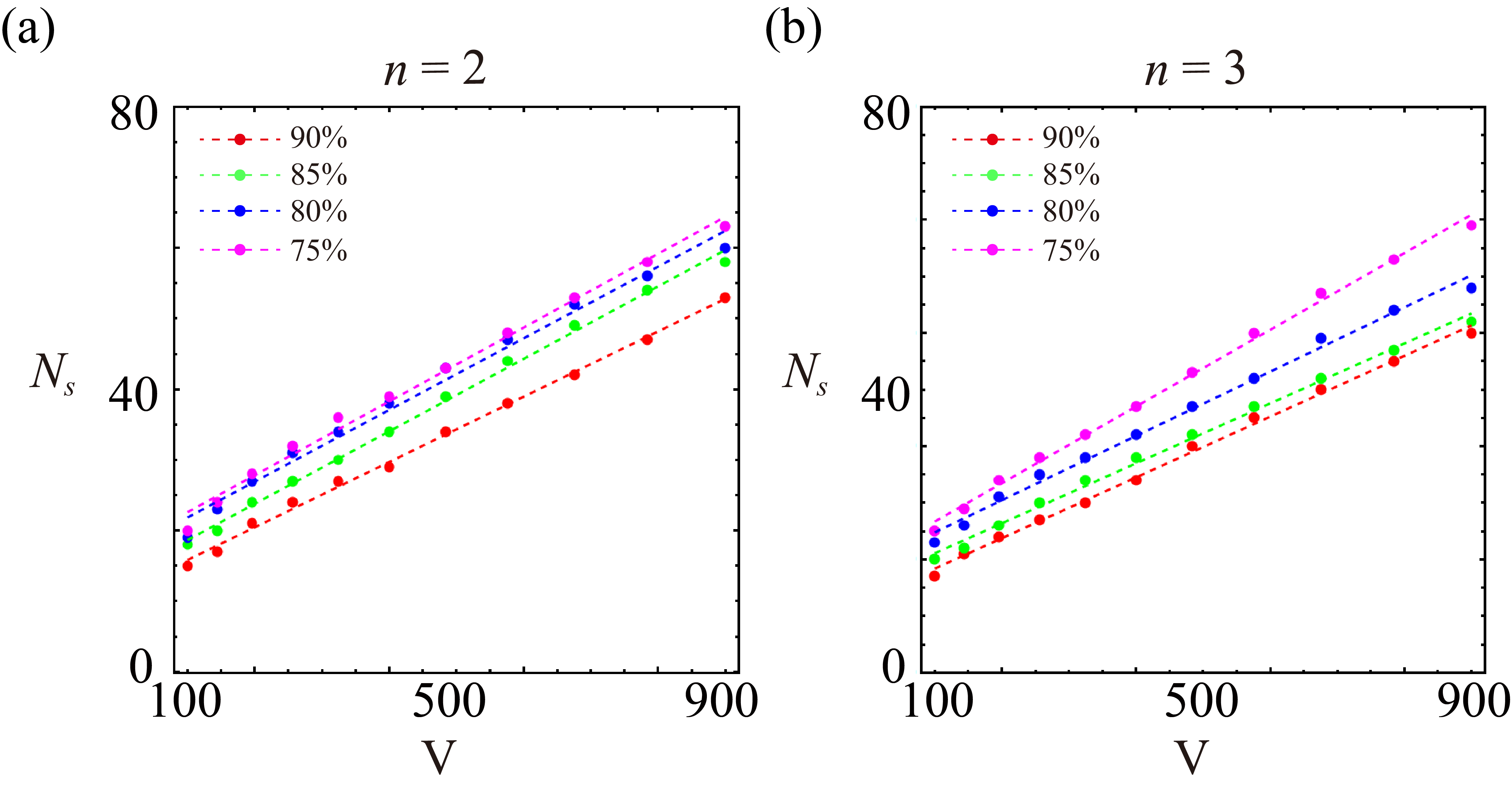}
\caption{\label{fig:figA6}Volume law of parallelogram geometry (for PhC) with different boundary thickness, where $n$ denotes the thickness of the boundary measured in unit cells. The markers correspond to different threshold values adopted for cases $\eta'_b/\eta'_t$.
}
\end{figure}

Additionally, in the Lieb PhC, we also verify the robustness of volume law with different thresholds $(75\%, 80\%, 85\%, 90\%)$ and boundary thicknesses ($n=2,3$), as shown in Fig.~\ref{fig:figA6}. The results show that $N_s$ and $V$ always maintain a linear relation, consistent with the conclusion under the condition of $80\%$ threshold and $n=2$ in the main text. \\

\begin{figure}[hbtp!]
\includegraphics[width=0.48\textwidth]{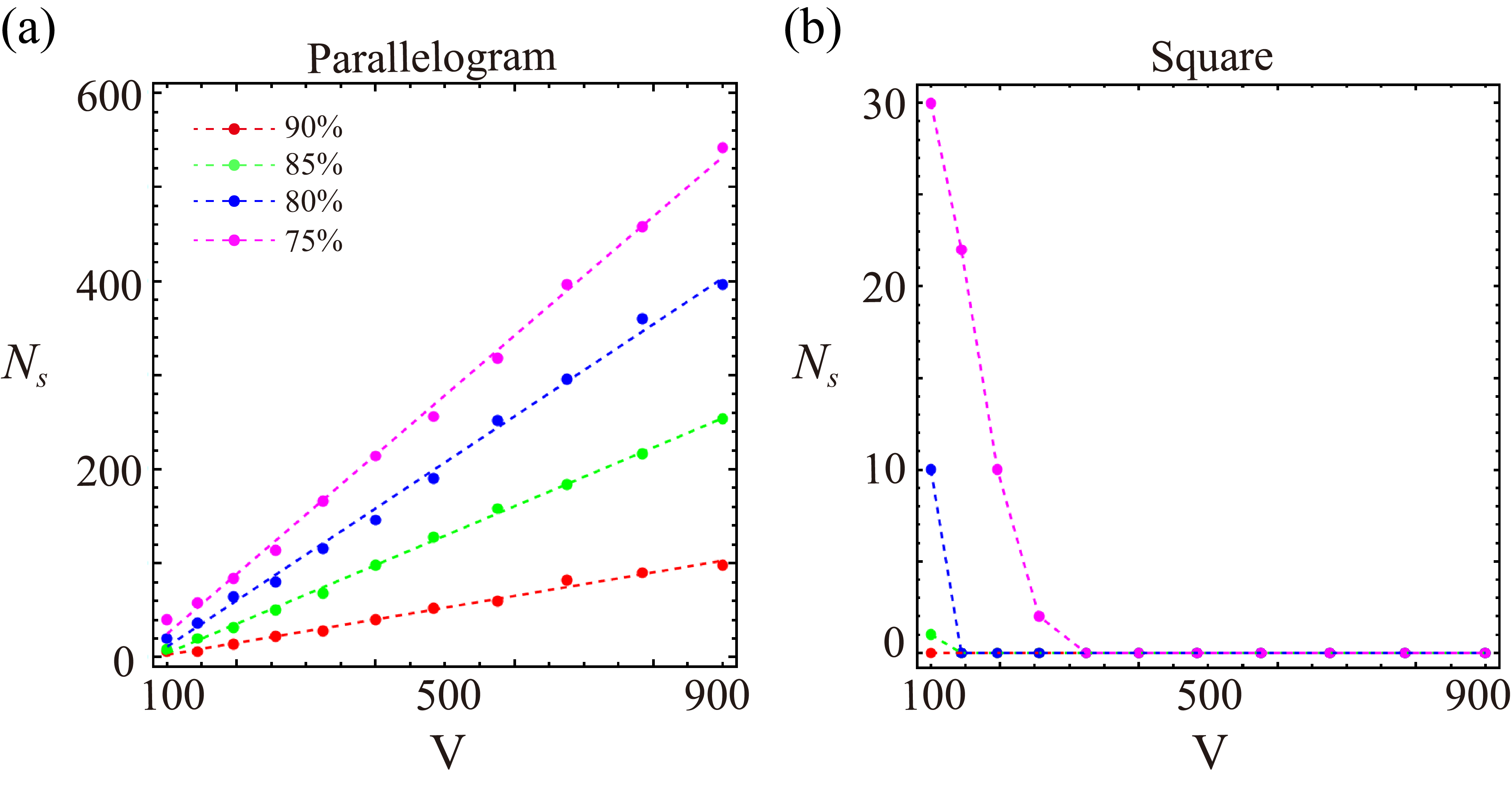}
\caption{\label{fig:figA7}Volume law of parallelogram geometry (for TBM) with boundary thickness $n=2$. The markers correspond to different threshold values adopted for cases $\eta_b/\eta_t$.
}
\end{figure}

In contrast, in the square structure, there is no linear relationship under all parameter combinations, as shown in Fig.~\ref{fig:figA7}(b). This indicates that the volume law of skin modes have good robustness, independent on moderate adjustment of the boundary localization threshold 
$(75\%-90\%)$ and boundary thickness $(n=2)$, and are only related to the boundary orientation and NH properties.

In summary, the verification results under different thresholds and boundary thicknesses further support the core conclusion of the geometry-dependent NHSE in this manuscript, proving the stable effect and providing flexible selection pool of parameters in future experimental works.

\section*{Acknowledgment}

Z.-K. X., and Y. L. thank Lan Zhi-Hao, Liu Zheng-Rong, and Chen Rui for useful discussion. The work is supported by National Natural Science Foundation of China (Grants Nos. 12074107, 62571212, U25D8012), Natural Science Foundation of Hubei Province of China (Grants Nos. 2024AFA038, 2022CFB553, 2022CFA012), Wuhan City Key R\&D Program (Grant No. 2025050602030069), Program of Outstanding Young and Middle-aged Scientific and Technological Innovation Team of Colleges and Universities in Hubei Province of China (Grant No. T2020001), and 2023 supplemental grant for 1A0702E004: Modern Optics.

\section*{Disclosures} The authors declare no conflicts of interest.

\section*{Data availability} Data underlying the results presented in this paper are not publicly available at this time but may be obtained from the authors upon reasonable request.



\ifCLASSOPTIONcaptionsoff
  \newpage
\fi



\bibliographystyle{IEEEtran}

%




\bibliography{reference_v2}

\end{document}